\documentclass{aa}
\usepackage{psfig}

\def\hii{H~II}

\def\hh{H$_2$}
\def\co{C$^{17}$O}
\def\cs{C$^{34}$S}
\def\water{H$_2$O}

\def\hhco{H$_2$CO}

\def\meth{CH$_3$OH}

\def\coo{CO$_2$}
\def\oo{O$_2$}
\def\hcch{C$_2$H$_2$}

\def\gtsim{{_>\atop{^\sim}}}
\def\ltsim{{_<\atop{^\sim}}}

\def\kms{km~s$^{-1}$}

\def\scm{cm$^{-2}$}
\def\ccm{cm$^{-3}$}
\def\mic{$\mu$m}
\def\klm{k$\lambda$}
\def\mjyb{mJy beam$^{-1}$}
\def\msol{M$_{\odot}$}
\def\lsol{L$_{\odot}$}

\newcommand\araa{{ARA\&A}}
\newcommand\aap{{A\&A}}

\newcommand\aaps{{A\&AS}}

\newcommand\apj{{ApJ}}
\newcommand\apjl{{ApJ}}
\newcommand\apjs{{ApJS}}
\newcommand\mnras{{MNRAS}}

\newcommand\nat{{Nature}}
\newcommand\pasp{{PASP}}

\begin{document}

\thesaurus{09.13.2; 08.03.2; 08.06.2}

\title{Abundance profiles of \meth\ and \hhco\ toward massive young
  stars as tests of gas-grain chemical models}

\titlerunning{\meth\ and \hhco\ toward massive YSOs}

\author{Floris F.S. van der Tak \inst{1} \and Ewine F. van Dishoeck
  \inst{1} \and Paola Caselli\inst{2} }
\institute{Sterrewacht, Postbus 9513, 2300 RA Leiden, The Netherlands
  \and Osservatorio Astrofisico di Arcetri, Largo E.\ Fermi 5, 50125
  Firenze, Italy}

\authorrunning{van der Tak et al.}
\offprints{Floris van der Tak \\ (vdtak@strw.leidenuniv.nl)}

\date{Received April 19, 2000 / Accepted July 6, 2000}

\maketitle

\begin{abstract}
  
  The chemistry of \meth\ and \hhco\ in thirteen regions of massive
  star formation is studied through single-dish and interferometer
  line observations at submillimeter wavelengths. Single-dish spectra
  at $241$~and $338$~GHz indicate that $T_{\rm rot} = 30-200$~K for
  \meth, but only $60-90$~K for \hhco. The tight correlation between
  $T_{\rm rot}$(\meth) and $T_{\rm ex}$(\hcch) from infrared
  absorption suggests a common origin of these species, presumably
  outgassing of icy grain mantles.  The \meth\ line widths are
  $3-5$~\kms, consistent with those found earlier for \co\ and \cs,
  except in GL~7009S and IRAS 20126, whose line shapes reveal \meth\ 
  in the outflows. This difference suggests that for low-luminosity
  objects, desorption of \meth-rich ice mantles is dominated by
  shocks, while radiation is more important around massive stars.
  
  The wealth of \meth\ and \hhco\ lines covering a large range of
  excitation conditions allows us to calculate radial abundance
  profiles, using the physical structures of the sources derived
  earlier from submillimeter continuum and CS line data. The data
  indicate three types of abundance profiles: flat profiles at
  \meth/\hh$\sim 10^{-9}$ for the coldest sources, profiles with a
  jump in its abundance from $\sim 10^{-9}$ to $\sim 10^{-7}$ for the
  warmer sources, and flat profiles at \meth/\hh\ $\sim$ few $10^{-8}$
  for the hot cores. The models are consistent with the $\approx 3''$
  size of the \meth\ $107$~GHz emission measured interferometrically.
  The location of the jump at $T\approx 100$~K suggests that it is due
  to evaporation of grain mantles, followed by destruction in
  gas-phase reactions in the hot core stage.  In contrast, the \hhco\ 
  data can be well fit with a constant abundance of a few $\times
  10^{-9}$ throughout the envelope, providing limits on its grain
  surface formation.  These results indicate that $T_{\rm rot}$
  (\meth) can be used as evolutionary indicator during the embedded
  phase of massive star formation, independent of source optical depth
  or orientation.
  
  Model calculations of gas-grain chemistry show that CO is primarily
  reduced (into \meth) at densities $n_{\rm H} \ltsim 10^4$~\ccm, and
  primarily oxidized (into \coo) at higher densities.  A temperature
  of $\approx 15$~K is required to keep sufficient CO and~H on the
  grain surface, but reactions may continue at higher temperatures if
  H~and O~atoms can be trapped inside the ice layer.  Assuming grain
  surface chemistry running at the accretion rate of CO, the observed
  abundances of solid CO, \coo\ and \meth\ constrain the density in
  the pre-protostellar phase to be $n_{\rm H} \gtsim$ a few
  $10^4$~\ccm, and the time spent in this phase to be $\ltsim
  10^5$~yr.  Ultraviolet photolysis and radiolysis by cosmic rays
  appear less efficient ice processing mechanisms in embedded regions;
  radiolysis also overproduces HCOOH and CH$_4$.
  

\end{abstract}

\section{Introduction}
\label{sec:intro}

\begin{figure*}[t]
  \begin{center}

\psfig{file=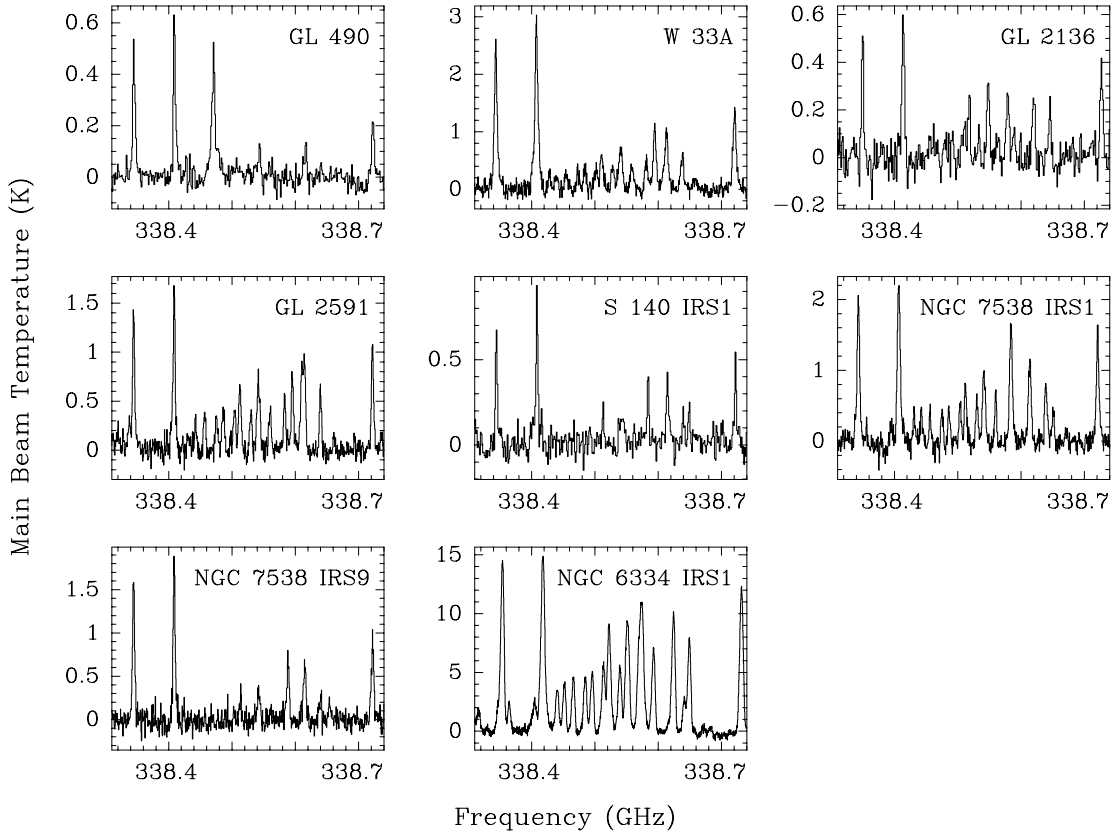,width=18cm,angle=-90}
    
    \caption{Spectra of the \meth\ $J=7_K\to 6_K$ transition at $338$~GHz, obtained with the JCMT.}
    \label{fig:jcmt}
  \end{center}
\end{figure*}

One of the early stages of massive star formation is the ``hot core''
phase (Kurtz et al.\ 2000\nocite{kurt00}, \cite{macd00}), which is
characterized by masses of $\sim 100$~\msol\ of molecular gas at
temperatures of $\gtsim 100$~K, leading to a rich line spectrum at
submillimeter wavelengths like that of the prototypical Orion hot
core. The chemical composition of hot cores is quite distinct for its
high abundances of fully hydrogenated, large carbon-bearing molecules
(\cite{ohis97}), as opposed to dark clouds which show predominantly
unsaturated species and molecular ions (\cite{evdgab98}). Saturated
carbon chains are not expected in large quantities by steady-state
gas-phase models, which led \cite{char92} and \cite{casel93} to
propose models where these molecules are made on dust grains, and
evaporate into the gas phase when the newly-formed star heats up its
surroundings. This leads to a short period ($\sim 10^4$~yr) when the
envelope of the young star is rich in complex organic molecules. Grain
surface reactions at low temperature lead to hydrogenation if atomic~H
is able to quantum tunnel through activation barriers as proposed by
\cite{tiel82}.  This model makes specific predictions for the
abundance ratios of CO, \hhco, \meth\ and their deuterated versions
which for a reasonable choice of parameters fit the abundances
observed in the Compact Ridge in Orion as shown by \cite{ctr97}.
However, it is unknown if other sources can be fit as well.
Alternatively, processing by ultraviolet light or energetic particles
may change the ice composition in the vicinity of young stars.

One key molecule to test models of hot core chemistry is methanol,
\meth.  At temperatures $\ltsim 100$~K, production of \meth\ in the
gas phase goes mainly via radiative association of CH$_3^+$ and
\water, an inefficient process yielding abundances of only $\sim
10^{-11}$ relative to \hh\ (Lee et al.\ 1996)\nocite{lee96}.
Observations of \meth\ toward dark and translucent clouds by
\cite{turn98} and \cite{takak98} yield abundances of $\sim 10^{-9}$,
which suggests that grain chemistry operates. Toward the Orion hot
core and compact ridge, \meth\ abundances of $\sim 10^{-7}$ have been
observed by \cite{blak87} and \cite{sutt95}, and even higher
abundances, $\sim 10^{-6}$, have been inferred from observations of
\meth\ masers (e.g., \cite{ment86}).  Methanol has also been observed
in the solid state at infrared wavelengths from the ground by, e.g.,
\cite{grim91} and \cite{all92} and recently by \cite{dart99b}, at
abundances up to $10^{-6}$.

To study the organic chemistry in warm ($30-200$~K) and dense ($10^4 -
10^7$~\ccm) circumstellar envelopes, this paper discusses observations
of \meth\ and \hhco\ lines at submillimeter wavelengths.  Although
gas-grain interactions are clearly important for methanol, it is
unknown how efficiently (if at all) surface reactions modify the
composition of the ices. In addition, it is still open what process
returns the molecules to the gas phase: thermal heating, shocks or
both? Finally, the ice layer must have formed at temperatures and
densities much lower than the present situation, and hence provides a
unique fossil record of the conditions in the molecular cloud prior to
star formation.

The sources are thirteen massive ($L=10^3-10^5$~\lsol) stars, which
are at an early stage of evolution and still embedded in
$10^2-10^3$~\msol\ of dust and molecular gas. A few of the sources are
hot cores by the definition of \cite{kurt00}, but most are in an even
younger phase where most of the envelope is still at low temperatures.
The physical structure of the soures has been studied by
\cite{fvdt00}, who developed detailed temperature and density
profiles. The wealth of \meth\ and \hhco\ lines, combined with the
physical models, allows the determination of abundance profiles
for \hhco\ and \meth, which demonstrate that \meth\ evaporates off
dust particles in the envelope on a $\sim 10^{4-5}$~yr time scale,
while \hhco\ is predominantly formed in the gas phase.  The excitation
and abundance of \meth\ are therefore useful evolutionary indicators
during the embedded stage of star formation.  Moreover, by comparing
the observed amounts of evaporated \meth\ and \coo\ to a model of
grain surface chemistry, we set limits on the density and the duration
of the pre-protostellar phase.

\section{Observations}
\label{sec:obs}

\begin{table*}[!t]
  \begin{center}
    \caption{Fluxes $\int T_{\rm MB}dV$ (K~\kms) and FWHM widths
      (\kms) of lines observed with the JCMT.}
    \label{tab:jcmt}

\begin{tabular}{rrrrrrrrrrrrr}
\hline
Line & Frequency & GL~490 & W~33A & GL & GL & S~140 & \multicolumn{2}{c}{NGC 7538} & NGC~6334 & GL & IRAS & W~28 \\
(K,A/E)$^a$
        & (MHz)  &        &     &2136 & 2591 & IRS1 &  IRS1 &  IRS9      & IRS1 & 7009S & 20126 & A2$^d$ \\
 & & & & & & & & & & \\ \hline
\multicolumn{13}{c}{$J=5\to 4$ band} \\
 & & & & & & & & & \\ \hline
 0 E   & 241700.2 & 0.8 & 5.3 & 0.7  & 2.0   & 1.5  & 4.8 & 3.0  &79.0 &  1.8 & 6.4 & 27.1  \\  
-1 E   & 241767.2 & 2.7 &12.3 & 0.9  & 3.3   & 3.0  & 8.5 & 6.7  &102. &  7.5 & 4.1 & 41.6  \\  
0 A$^+$& 241791.4 & 3.0 &13.8 & 1.1  & 4.0   & 3.5  & 9.7 & 7.6  &101. &  8.5 & 8.4 & 39.5  \\  
4 A    & 241806.5 &$<0.2$& 1.2 & 0.2  & 0.6  &$<0.2$& 1.2 &$<0.2$&31.6 &$<0.2$ & 2.2 &  3.0  \\  
-4 E   & 241813.2 &$<0.2$& 0.7 & 0.2  & 0.4  &$<0.2$& 0.8 &$<0.2$&21.4 & $<0.2$ & 1.1 &  0.9  \\  
4 E    & 241829.6 &$<0.2$& 0.8 &$<0.2$& 0.3  &$<0.2$& 1.4 &$<0.2$&25.5 & $<0.2$ & 0.8 &  2.6  \\  
3 A    & 241832.9 &$<0.2$& 1.9 &$<0.2$& 1.3  & 0.5  & 1.6 & 1.2  &36.0 & $<0.2$ & 3.2 & 11.5  \\  
3 E /2 A$^-$& 241843.0 & 0.3 & 2.2 & 0.4  & 1.3  &0.5   & 2.4 & 0.9  & 55.4 &$<0.2$ & 4.1 & 12.0   \\  
-3 E   & 241852.3 &$<0.2$& 0.9 &$<0.2$& 0.6  &$<0.2$& 0.9 &$<0.2$& 28.0 &$<0.2$ & 1.5 &  3.6   \\  
1 E    & 241879.0 & 0.5 & 3.7 & 1.0  & 1.6  &1.2   & 3.5 & 2.0  & 57.6 &$<0.2$ & 4.4 & 20.6   \\  
2 A$^+$& 241887.7 & 0.2 & 2.0 & 0.5  & 1.0  &0.3   & 1.5 & 0.6  & 38.9 &$<0.2$ & 2.4 &  8.2   \\  
$\mp2$E& 241904.4 & 0.8 & 4.8 & 1.0  & 2.3  &1.5   & 5.0 & 2.8  & 78.3 & 0.9 & 5.8 & 26.1   \\  
 & & & & & & & & & \\ \hline
\multicolumn{13}{c}{$J=7\to 6$ band} \\
 & & & & & & & & & \\ \hline
-1 E   & 338344.6 & 2.1  &14.1 & 2.1  & 5.2  & 2.1  & 8.8 & 6.0  &86.8 & ... & ... & 27.6 \\
0 A$^+$~$^b$
       & 338408.6 & 2.2  &17.9 & 2.8  & 6.2  & 2.8  &11.3 & 6.6  &128.6 & ... & ... & 25.7 \\
-6 E   & 338430.9 &$<0.3$& 1.2 &$<0.5$&$<0.5$&$<0.5$& 1.4 &$<0.5$ &23.6 & ... & ... & $<3$ \\
6 A    & 338442.3 &$<0.3$& 1.4 &$<0.5$& 1.1  &$<0.5$& 1.5 &$<0.5$ &23.7 & ... & ... & $<3$ \\
-5 E   & 338456.5 &$<0.3$& 1.6 &$<0.5$& 1.5  &$<0.5$& 1.2 &$<0.5$ &26.2 & ... & ... & $<3$ \\
5 E    & 338475.2 &$<0.3$& 1.6 &$<0.5$& 1.5  &$<0.5$& 1.5 &$<0.5$ &29.5 & ... & ... & $<3$ \\
5 A    & 338486.3 &$<0.3$& 1.9 &$<0.5$& 1.9  &$<0.5$& 1.6 &$<0.5$ &31.4 & ... & ... & $<3$ \\
-4 E   & 338504.0 &$<0.3$& 2.2 &$<0.5$& 1.8  &$<0.5$& 2.0 &$<0.5$ &35.2 & ... & ... & $<3$ \\
4A / 2A$^-$ &338512.7  &$<0.3$& 3.5 & 1.1  & 2.8  & 0.6  & 3.1 & 0.8   &59.4 & ... & ... & 11.6 \\
4 E    & 338530.2 &$<0.3$& 2.3 &$<0.5$& 1.5  &$<0.5$& 2.3 &$<0.5$ &33.3 & ... & ... & $<3$ \\
3 A    & 338541.9 & 0.5  & 5.2 & 1.5  & 3.7  & 1.3  & 4.9 & 1.6   &65.7 & ... & ... & 13.5 \\
-3 E   & 338559.9 &$<0.3$& 2.6 &$<0.5$& 1.8  &$<0.5$& 2.2 & 0.4   &$<102$ & ... & ... & $<3$ \\
3 E$^c$& 338583.1 &$<0.3$& 9.2 &$<0.5$& 2.1  & 1.3  & 8.1 & 2.9   &39.4 & ... & ... & 3.8 \\
1 E    & 338615.0 & 0.6  & 7.2 &  1.3 & 2.7  & 1.7  & 3.1 & 2.8   &57.3 & ... & ... & 42.5 \\
2 A$^+$& 338639.9 & 0.2  & 3.7 &  0.2 & 2.4  & 0.6  & 3.6 & 1.1   &41.1 & ... & ... & 11.4 \\
$\mp2$E& 338722.5 & 1.0  & 8.8 &  2.0 & 4.0  & 1.5  & 6.9 & 3.9   &74.0 & ... & ... & 24.9 \\ 
 & & & & & & & & & \\ \hline
\multicolumn{13}{c}{Line Width } \\
 & & & & & & & & & \\ \hline
 & & 3.4   & 5.2  & 3.0  & 3.5  & 2.7  & 3.9  & 3.6  & 5.7 & 6.9  & 7.5 & 6.2   \\ 
 & &$\pm 0.7$&$\pm 1.1$&$\pm 1.3$&$\pm 0.4$&$\pm 0.2$&$\pm 0.6$&$\pm 0.7$&$\pm 0.5$&$\pm 1.3$&$\pm 1.9$&$\pm 0.9$\\ \hline

\end{tabular}
  \end{center}

$^a$ When no superscript is given for A-type methanol, the A$^+$ and A$^-$ lines are 
blended.

$^b$ Blend with K=6 E at 338404.5 MHz, assumed to equal K=-6 E.

$^c$ Possible contribution from K=1 A$^-$ at 341415.6 MHz.

$^d$ This source is also known as G~5.89-0.39; $338.5$~GHz data are from Thompson
                  \& Macdonald 1999.
 
\end{table*}

\begin{figure}[t]
  \begin{center}
    
\psfig{file=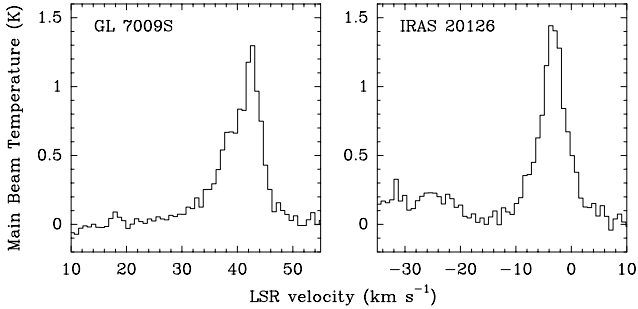,width=9cm,angle=-90}

    \caption{Line profiles of \meth\ $5_{-0}\to4_{-0}$~A$^+$ toward
      GL~7009S and IRAS 20126. The outflow wing is most prominent in
      GL~7009S. The feature at $-25$~\kms\ in IRAS 20126 is the
      $K=4$~A line.}
    \label{fig:lprof}

  \end{center}
\end{figure}

\subsection{Single-dish observations}
\label{sec:obs_jcmt}

\begin{figure}[t]
  \begin{center}
    
\psfig{file=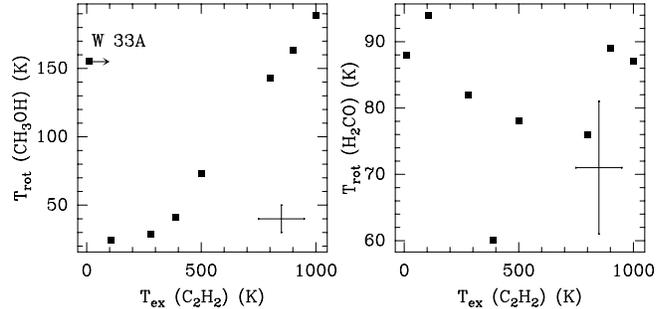,width=9cm,angle=-90}

    \caption{Rotation temperatures of \meth\ ({\it left})
      and \hhco\ ({\it right}) versus excitation temperatures of
      \hcch\ measured in infrared absorption. Crosses in the bottom
      right denote typical error margins. }
    \label{fig:rot}

  \end{center}
\end{figure}

Spectroscopy of the $J=5\to4$ and $7\to6$ bands of \meth\ near
$242$~and $338$~GHz was performed with the 15-m James Clerk Maxwell
Telescope (JCMT)\footnote{The James Clerk Maxwell Telescope is
  operated by the Joint Astronomy Centre, on behalf of the Particle
  Physics and Astronomy Research Council of the United Kingdom, the
  Netherlands Organization for Scientific Research and the National
  Research Council of Canada.} on Mauna Kea, Hawaii in May and October
of 1995. The beam size (FWHM) and main beam efficiency of the antenna
were $18''$ and $69$\% at $242$~GHz and $14''$ and $58$\% at $338$~GHz.
The frontends were the receivers A2 and B3i; the backend was the
Digital Autocorrelation Spectrometer, covering $500$~MHz instantaneous
bandwidth. Pointing was checked every 2 hours during the observing and
was always found to be within $5''$. To subtract the atmospheric and
instrumental background, a reference position $180''$~East was
observed. Integration times are 30--40 minutes for each frequency
setting, resulting in rms noise levels in $T_{\rm mb}$ per $625$~kHz
channel of $\approx 30$~mK at 242~GHz to $\approx 50$~mK at 338~GHz.
Although the absolute calibration is only correct to $\approx 30$\%,
the relative strength of lines within either frequency setting is much
more accurate.

In addition to these \meth\ lines, this paper discusses observations
of lines of \hhco, obtained in a similar manner, which were presented
in \cite{fvdt00}.  We will also use data on both molecules for W~3
IRS5 and W~3 (\water) from the survey by \cite{helm97}, and \meth\
$J=7\to6$ data from \cite{thomp99}.

\subsection{Interferometric observations}
\label{sec:obs_ovro}

Maps of the $J_K=3_1\to4_0$~A$^+$ (107013.852~MHz; $E_u = 28.01$~K)
and $11_{-1}\to10_{-2}$~E (104300.461~MHz; $E_u = 157.24$~K) lines of
\meth\ were obtained for four sources in 1995-1999 with the millimeter
interferometer of the Owens Valley Radio Observatory
(OVRO)\footnote{The Owens Valley Millimeter Array is operated by the
  California Institute of Technology under funding from the U.S.
  National Science Foundation (AST-9981546).}.  This instrument
consists of six 10.4~m antennas on North-South and East-West
baselines, and a detailed technical description is given in
\cite{padin91}. For the sources GL~2591, NGC~7538 IRS1 and NGC~7538
IRS9, data were taken in the compact and extended configurations,
while for the Southern source W~33A, a hybrid configuration with long
North-South but short East-West baselines was also used to improve the
beam shape.  Antenna spacings range from the shadowing limit out to
90~\klm, corresponding to spatial frequencies of
$0.018-0.44$~arcsec$^{-1}$. The observations were carried out
simultaneously with continuum observations, which have been presented
in van der Tak et al.~(1999, 2000). We refer the reader to these
papers for further observational details.

\section{Results}
\label{sec:res}

\begin{table*}
    \caption{Beam-averaged column densities and excitation
      temperatures of \meth\ and \hhco,
      and abundances inferred for a power law envelope using spherical Monte Carlo models.}
  \begin{center}
    \begin{tabular}{lcccccc}

\hline
\\
Source & $N$(\meth)  & $N$(\hhco) & $T$(\meth)  & $T$(\hhco) & \meth/\hh\ & \hhco/\hh$^{(c)}$ \\  
 & \scm\ & \scm\ & K & K & $10^{-9}$ & $10^{-9}$ \\
 & & \\ \hline
W 3 IRS5      & $1.6(14)$ & $7.8(13)$ & $73$ & $78$    & $0.4$   & 3 \\             
GL 490        & $3.6(14)$ & $4.3(13)$ & $24$ & $94$    & $1.0$   & 1 \\             
W 33A         & $2.0(15)$ & $1.2(14)$ & $155$ & $88$   & $3.1$ / $90$$^{(b)}$ & 4 \\
GL 2136       & $4.4(14)$ & $4.5(13)$ & $143$ & $76$   & $0.9$   & 8 \\              
GL 7009S$^{(a)}$                                       
              & $3.9(15)$ & $...$     & $8$ & $...$    & $0.7$   & ... \\              
GL 2591       & $1.2(15)$ & $8.0(13)$ & $163$ & $89$   & $2.6$ / $80$$^{(b)}$ & 4 \\
S 140 IRS1    & $2.5(14)$ & $8.0(13)$ & $41$ & $60$    & $1.2$   & 5 \\              
NGC 7538 IRS1 & $2.2(15)$ & $1.9(14)$ & $189$ & $87$   & $2.0$ / $60$$^{(b)}$ & 10 \\
NGC 7538 IRS9 & $6.5(14)$ & $1.0(14)$ & $29$ & $82$    & $2.3$   & 10 \\              
W 3 (\water)  & $7.5(15)$ & $5.0(14)$ & $203$ &$181$   & $5.9$   & 3 \\              
NGC 6334 IRS1 & $3.8(16)$ & $1.6(15)$ & $213$ &$193$   & $24.$   & 7 \\             
IRAS 20126$^{(a)}$                                             
              & $2.2(15)$ & $...$ & $139$ & $...$ & $2.6$   & ... \\ 
W~28 A2        & $4.2(15)$ & $...$ & $43$  & $...$ & $12.$   & ... \\             

\hline

    \end{tabular}
    \label{t:trot}
  \end{center}

{\scriptsize(a):} Only \meth\ $J=5\to4$ observed.

{\scriptsize(b):} Abundances in cold ($T<90$~K) and in warm ($T>90$~K) gas.

{\scriptsize(c):} From van der Tak et al. 2000.

\end{table*}

\subsection{Submillimeter spectroscopic results}
\label{sec:res_jcmt}

\begin{figure*}[t]
  \begin{center}
    
\psfig{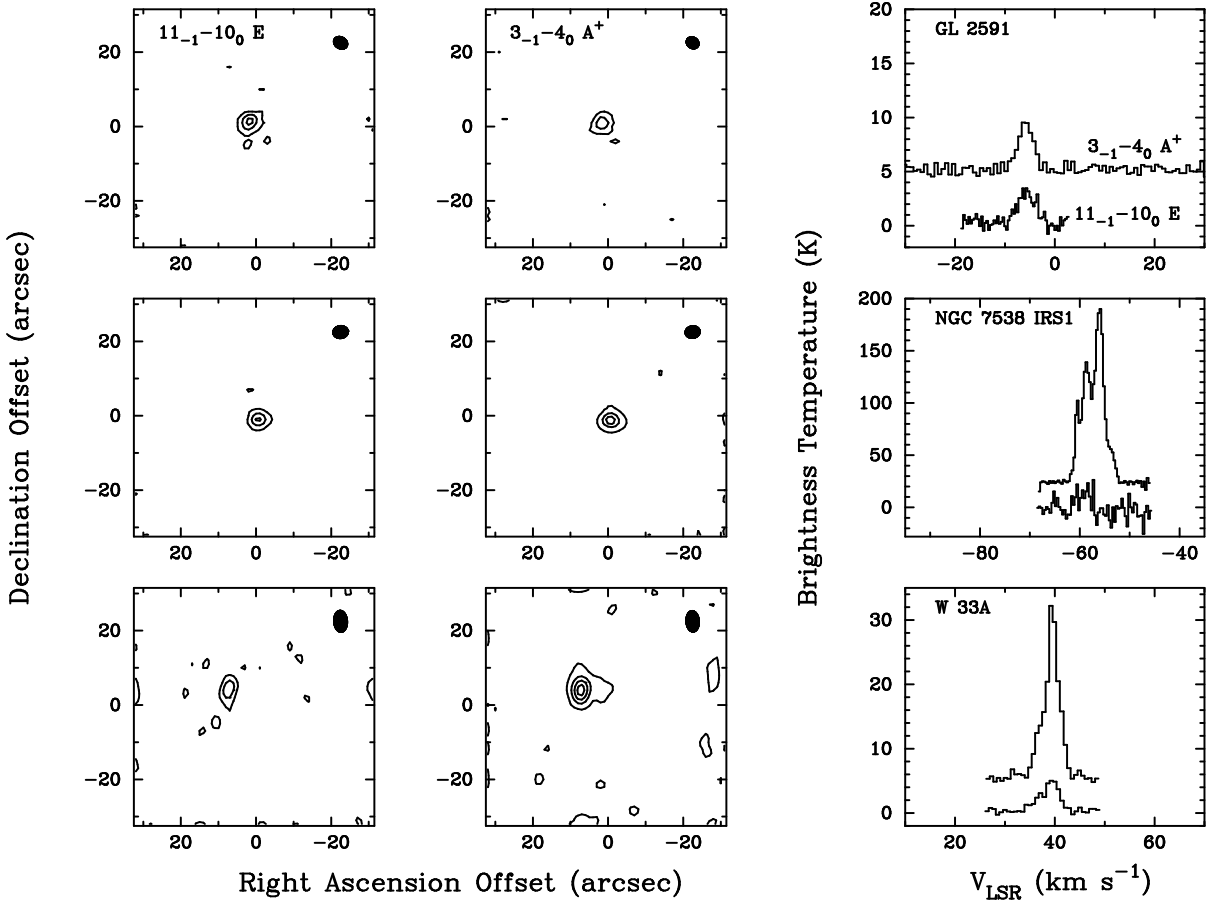}

    \caption{Maps and spectra of \meth\ line emission, made with
      the OVRO array. Contour levels for GL~2591 (top row) are: 60, 180,
      300~\mjyb\ for both images; for NGC~7538 IRS1 (middle row): 300, 900,
      1500~\mjyb\ (left) and 400, 2400 and 4400~\mjyb\ (right); and
      for W~33A (bottom row): 100, 200~\mjyb\ (left) and 100, 300,
      500, 700~\mjyb\ (right). In the spectrum of NGC~7538 IRS1, the
      brightness of the $11_{-1}-10_{0}$~E line has been increased by
      a factor of ~4.}
    \label{fig:ovro}

  \end{center}
\end{figure*}

Figure~\ref{fig:jcmt} presents the calibrated JCMT spectra, reduced
using the IRAM CLASS package. All lines are due to either A- or E-type
\meth, while some features are blends of lines.  After subtraction of
a linear baseline, Gaussian profiles were fitted to extract line
parameters. The free parameters in these fits were the line flux $\int
T_{\rm mb} dV$, the FWHM line width and the central velocity. The
retrieved line fluxes and widths are listed in Table~\ref{tab:jcmt},
while the central velocities are consistent with those found for
\co\ and \cs\ in \cite{fvdt00}. 

For most sources, the line widths are also consistent with those found
for \co\ and \cs. However, for the two lowest-luminosity sources,
GL~7009S and IRAS~20126, the methanol lines are much broader than
those of \co\ and \cs. The \meth\ line profiles toward these sources
(Fig.~\ref{fig:lprof}) suggest an origin in their molecular outflows,
especially in GL~7009S. The methanol emission toward young low-mass
stars such as L~1157 (\cite{bach95}) and NGC~1333 IRAS~4A
(\cite{blak95}) is also dominated by their outflows. The other sources
studied here possess outflows, but those do not seem to be important
for the \meth\ emission. As we will show below, the observed gas-phase
\meth\ is mostly due to evaporation of grain mantles. The \meth\ line
profiles thus suggest that around low-mass stars, the desorption of
icy grain mantles is dominated by shocks, while radiative heating
dominates for high-mass stars.

The sources exhibit clear variations in their methanol spectra, both
in absolute line strength and in the relative strength between lines.
Variations in absolute line strength can be due to differences in
molecular abundances, the size of the emitting region or distance,
while the line ratios reflect differences in the temperature and
density of the emitting gas.  Before modeling the observations in
detail, we perform a quick-look examination using rotation diagrams,
as in \cite{blak87} and \cite{helm94}.  This analysis assumes that the
emission is optically thin and that the population of the molecular
energy levels can be described by a single temperature, the
``rotational temperature''.  Since the envelopes of these stars are
known to have temperature and density gradients, the rotation
temperature gives information where the molecule is preferentially
located in the envelope. The nondetection of $^{13}$\meth\ lines
limits the optical depth of the lines to $\ltsim 3$, except in the
case of NGC~6334 IRS1, where the strongest lines are probably
saturated.

Table~\ref{t:trot} presents the results of the rotation diagram
analysis, which are averages over the JCMT beam. Formal errors from
the least-square fit to the data and calibration errors combine into
uncertainties of $20$\% in both parameters. The results for
IRAS~20126, W~3 (\water) and NGC~6334 IRS1 are more uncertain, to
$\approx 50$\%, because the derived $T_{\rm rot}$ is close to the
highest observed energy level. The OVRO maps presented in
\S~\ref{sec:res_ovro} indicate that the \meth\ column densities are
beam diluted by a factor of $\sim 25$; this information is not
available for \hhco. Although the column densities of the two species
are roughly correlated, with $N$(\meth) $\sim 10 \times N$(\hhco),
their excitation temperatures behave differently.  While $T_{\rm
  rot}$(\hhco) is rather constant from source to source, $60-90$~K,
except toward the hot cores W~3 (\water) and NGC~6334 IRS1, the \meth\ 
excitation temperatures span the full range $10-200$~K.

The spectral line survey of \cite{thomp99} toward W~28~A2 includes
  the \meth\ $J=7\to6$ band, and their analysis gives $T_{\rm
    rot}=49$~K and $N=3.8\times 10^{15}$~\scm, in good agreement with
  our results from the $J=5\to4$ band. Our values of $N$(\hhco) toward
  W~3 (\water) and NGC~6334 IRS1 are a factor of $\sim$10 higher than
  those found by \cite{mang93}, probably due to the smaller JCMT beam
  size. 

Figure~\ref{fig:rot} compares the rotational temperatures of \meth\ 
and \hhco\ to the \hcch\ excitation temperatures measured by
\cite{fred00} in absorption at $13.7$~\mic\ for the same lines of
sight.  Lacking a dipole moment, \hcch\ is a clean probe of
temperature. The tight correlation between the \meth\ and \hcch\ 
excitation temperatures implies that the two species trace the same
gas, and that source orientation and optical depth effects do not
influence the infrared data. The only exception is W~33A: this massive
source becomes optically thick at $14$~\mic\ before the warm gas is
reached.  In contrast, the \hhco\ temperature is not correlated with
that of \hcch: omitting W~33A, the correlation coefficient is $7.8$\%,
versus $98.3$\% for \meth.  This difference indicates that in these
sources, \hhco\ is not chemically related to \meth\ and \hcch, unlike
in low-mass objects, as the high \hhco\ abundance in the L~1157
outflow (\cite{bach97}) and the detection of HDCO and D$_2$CO in IRAS
16293 (\cite{evd95}; \cite{cecc98}) indicate.

\subsection{Interferometric maps of \meth\ line emission}
\label{sec:res_ovro}

\begin{table*}
    \caption{Parameters of line emission observed with OVRO}
  \begin{center}
    \begin{tabular}{llrrrcc}

\hline
\\
Source & Line & LSR Velocity & $\Delta V$ (FWHM) & Peak $T_B$ & Source Size & Beam Size \\ 
 & & \kms\ & \kms\ & K & arcsec & arcsec \\
\hline
\\
GL 2591 & $3_1\to4_0$~A$^+$ & -5.8 & 3.5 & 4.3 & $3.4 \times 2.7$ & $3.4 \times 2.9$ \\
    & $11_{-1}\to10_{-2}$~E & -5.7 & 4.0 & 2.8 & $3.1 \times 2.3$ & $3.8 \times 3.0$ \\
W~33A   & $3_1\to4_0$~A$^+$ & 39.5 & 3.5 & 23.3 & $3.9 \times 3.0$ & $5.6 \times 3.4$ \\
    & $11_{-1}\to10_{-2}$~E & 39.1 & 4.3 & 4.4 & $6.4 \times 2.2$ & $5.7 \times 3.5$ \\
NGC 7538 IRS1 & $3_1\to4_0$~A$^+$ & -60.5 & 1.4 & 63.6 & -- & $3.9 \times 3.2$ \\
              &                   & -58.6 & 1.8 & 108. & -- & $3.9 \times 3.2$ \\
              &                   & -56.1 & 1.9 & 164. & -- & $3.9 \times 3.2$ \\
              &                   & -53.9 & 2.5 & 29.2 & -- & $3.9 \times 3.2$ \\
          & $11_{-1}\to10_{-2}$~E & -59.0 & 2.8 & 4.2 & $1.8\times 0.7$ & $4.0 \times 3.3$ \\
\hline 
\\
    \end{tabular}
    \label{tab:ovro}
  \end{center}
\end{table*}

Figure~\ref{fig:ovro} presents the OVRO observations, as maps of the
emission integrated over the line profile. These maps were obtained by
a gridding and fast Fourier transform of the $(u,v)$-data with natural
weighting, and deconvolution with the CLEAN algorithm. The
self-calibration solutions obtained for the continuum data were used
to suppress phase noise introduced by the atmosphere. Also shown are
spectra at the image maxima, extracted from image cubes at full
spectral resolution and coverage. In these maps and spectra, the
continuum emission has not been subtracted. Only in the case of
NGC~7538 IRS1, the continuum brightness is comparable to the line
brightness; for the other sources, the continuum can be neglected.

The maps show single, compact emitting sources, which coincide within
the errors with their counterparts in $86-230$~GHz continuum emission
(van der Tak et al. 1999, 2000)\nocite{fvdt99,fvdt00}. No \meth\ line
emission was detected toward NGC~7538 IRS9 to the $1\sigma$ noise
level of $\approx 1$~K.  In the case of W~33A, a binary source at
these wavelengths, the \meth\ emission is associated with the source
MM1 from \cite{fvdt00} and with the infrared source. Although the
\meth\ emission is seen to be slightly extended toward the West, no
emission was detected at the position of the other continuum source,
MM2.

Table~\ref{tab:ovro} lists various parameters of the lines observed
with OVRO. Deconvolved source sizes were obtained by fitting 
two-dimensional Gaussian profiles to the integrated emission maps;
other parameters by fitting Gaussians to the presented spectra.
The continuum emission was subtracted from the data before this
fitting procedure.

The measured central velocities and line widths are consistent with
the values measured with the JCMT. To check if the two telescopes
indeed trace the same gas, we have calculated the line strengths
expected from the JCMT rotation diagrams.  For the sources where
\meth\ is detected with OVRO, we predict a brightness of $15-30$~K in
the \meth\ $107$~GHz line and of $10-15$~K in the $104$~GHz line, in
good agreement with the data. For NGC~7538 IRS9, which has a low
\meth\ rotation temperature, we expect $T_B \approx 60$~K in the
$107$~GHz line. The lack of \meth\ emission in the OVRO data suggests
that there is a gradient in $T_{\rm ex}$ from $\approx 30$~K on a
$15''$ scale (traced by the JCMT but resolved out by OVRO) to $\approx
100$~K within the $3''$ OVRO synthesized beam.  For the other sources,
the bulk of the envelope may already be heated to $\approx 100$~K and
up.


\subsection{Maser emission}
\label{sec:maser}

While the $V_{\rm LSR}=-53.9$~\kms\ component in the $107$~GHz line
profile observed toward NGC~7538 IRS1 with OVRO has a brightness and
a width consistent with the JCMT data of this source, the brightness
of the other velocity components is a factor of $2-5$ higher than
expected from the JCMT data. This difference suggests that the
observed emission is amplified by a maser process. This possibility is
supported by the line profile with three narrow velocity peaks
(Figure~\ref{fig:ovro}).  Maser emission has been observed toward
this source in several lower-frequency methanol lines, by
\cite{wils84} and \cite{batr87}, at velocities consistent with the two
strongest components reported here.  Using VLBI, \cite{mini98}
resolved the maser emission in the $5_{1}\to 6_{0}$~A$^+$ ($6.67$~GHz)
and $2_{0}\to 3_{-1}$~E ($12.2$~GHz) lines into eleven spots, with
sizes of $<10^{-3}$ arc seconds each.  The spots at velocities $-57\to
-62$~\kms\ lie in an outflow, but the kinematics of brightest masers,
those at $\approx -56$~\kms, are consistent with an origin in a disk.
These two lines as well as the $107$~GHz line are transitions to the
$K-$ground state (so-called Class~II masers), which require a strong
continuum radiation field to be pumped (\cite{crag92}). In the case of
NGC~7538 IRS1, this radiation is provided by the H~II region,
consistent with the fact that among the sources studied here with
OVRO, only NGC~7538 IRS1 displays maser emission, which is also the
source with the strongest \hii\ region.  For the $1_{10}\to1_{11}$
line of \hhco, where \cite{rots81} observed maser action at similar
velocities, the same pump mechanism operates, as demonstrated by
\cite{bolan81}.

The brightness ratio of the $6.67$ to $12.2$~GHz methanol maser lines
measured by \cite{mini98} is $\approx 10$. Comparing this number with
the models by \cite{sobol97}, we find $N$(\meth)$\sim 10^{16}$~\scm.
This value is similar to $N$(\hhco) from \cite{bolan81}, which is a
lower limit because \cite{rots81} did not resolve the emission
spatially.  The ratio $N$(\hhco)/$N$(\meth) for the disk of NGC~7538
IRS1 is therefore $\gtsim 1$, which is significantly larger than in
the extended envelopes.

\section{Abundance profiles of \meth\ and \hhco}
\label{sec:models}

\begin{figure}[p]
  \begin{center}
    
\psfig{file=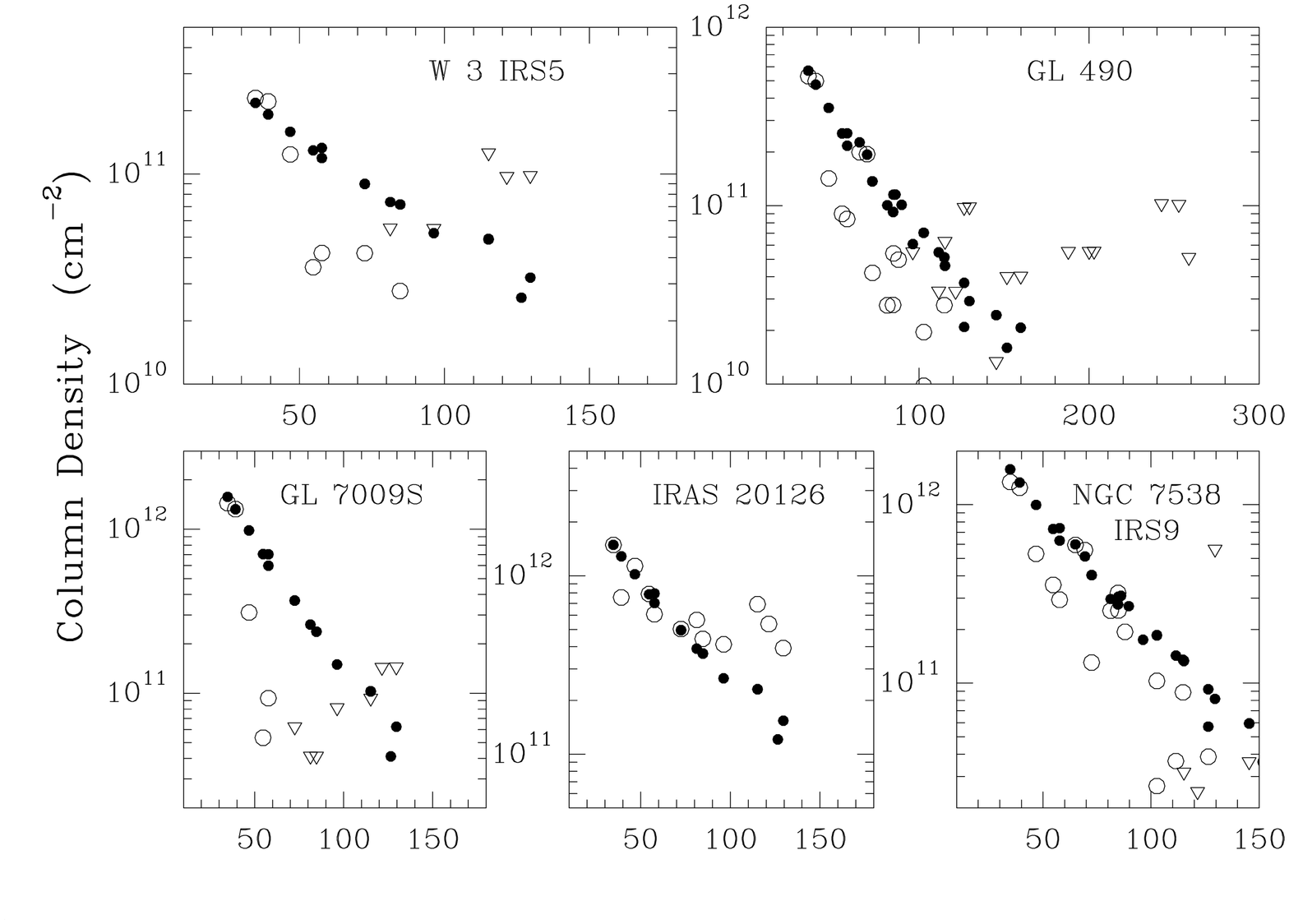,width=9cm,angle=-90}

\vfill

\psfig{file=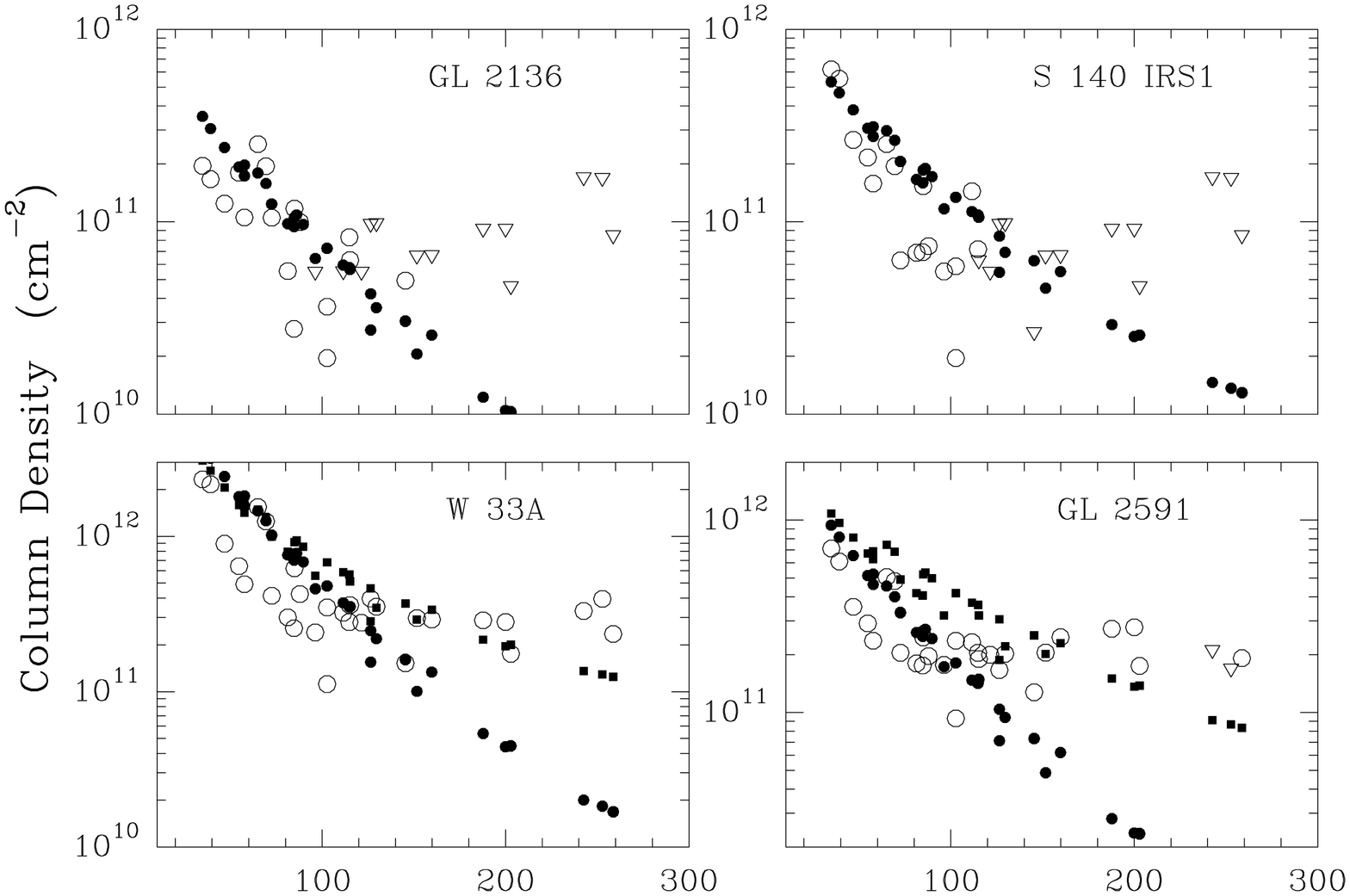,width=9cm,angle=-90}

\vfill

\psfig{file=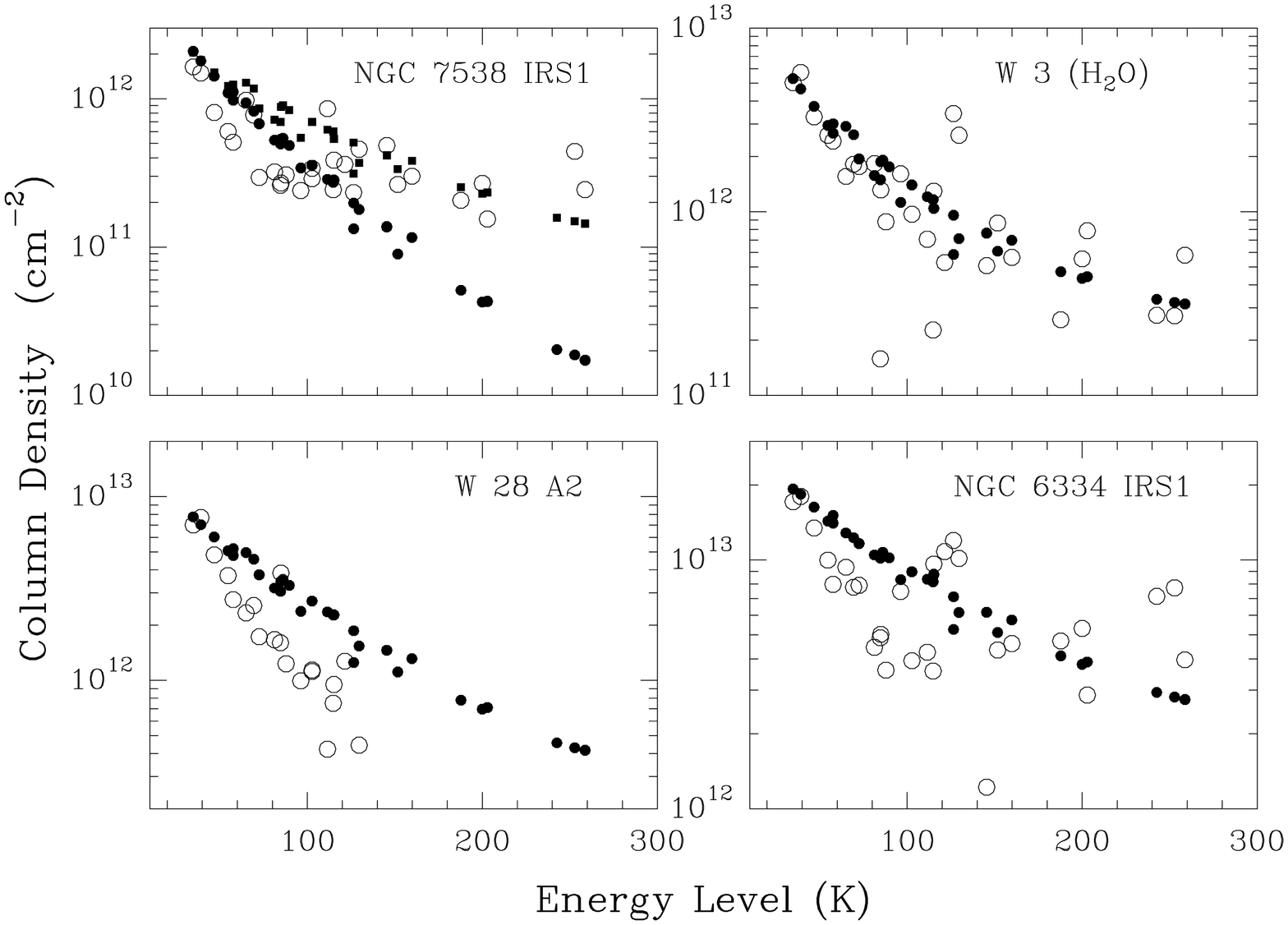,width=9cm,angle=-90}

\vfill

    \caption{Results of Monte Carlo models for \meth, with `cold'
      sources at the top, `jump' sources in the middle and `hot cores'
      at the bottom. Filled symbols are line strengths predicted by the best-fit
      constant-abundance models (circles) and jump models (squares)
      for W~33A, GL~2591 and NGC~7538 IRS1.  Open symbols are JCMT data,
      with circles indicating detections and triangles upper limits.
      The data on W~28~A2 include those of Thompson \& Macdonald 1999.}
    \label{fig:const_x}

\vfill

  \end{center}
\end{figure}

In this section, we proceed to interpret the \meth\ and \hhco\ 
observations in terms of the detailed physical models developed by
\cite{fvdt00}. These models use a power law density structure, $n=n_0
(r/r_0)^{-\alpha}$. Specific values for the density gradient $\alpha$,
the size scale $r_0$ and the density scale $n_0$ for every source can
be found in \cite{fvdt00}. These values, as well as temperature
profiles, have been derived from dust continuum and CS, \cs\ and
\co\ line observations at submillimeter wavelengths.

\subsection{Constant-abundance models}
\label{sec:mod_x0}

Using the density and temperature profiles, we have first attempted to
model the data using constant abundances of \hhco\ and \meth\ 
throughout the envelopes. This assumption is motivated by the fact
that in models of pure gas-phase chemistry, the abundances of \hhco\ 
and \meth\ do not change much in the range $20-100$~K. Molecular
excitation and radiative transfer are solved simultaneously with a
computer program based on the Monte Carlo method, written by
\cite{hst00}. The emission in all observed lines is calculated,
integrated over velocity, and convolved to the appropriate telescope
beam. Comparison to observations proceeds with the $\chi^2$ statistic
described in \cite{fvdt00}.

The collisional rate coefficients for \meth\ have been provided by
M. Walmsley (1999, priv. comm.), and are based on the experiments by
\cite{lees74}. A detailed description of the coefficients is given by
\cite{turn98}. Following \cite{john92}, we have set the $\Delta K=3$ 
rates to $10$\% of the $\Delta K=1$ values.

The results of the model calculations are shown in
Fig.~\ref{fig:const_x}, and the derived best-fit abundances are given
in column~6 of Table~\ref{t:trot}. In the figure, the observed
and modeled line fluxes have been converted to column densities
following \cite{helm94}. Three types of sources can be distinguished.
Most sources can be fitted with \meth/\hh$\sim 10^{-9}$, similar to
the values found for dark and translucent clouds (Turner 1998,
Takakuwa et al.\ 1998)\nocite{turn98,takak98}, while the ``hot
core''-type sources W~3(\water), NGC~6334 IRS1 and W~28 A2 require
abundances of a few times $10^{-8}$. However, for three sources, no
single methanol abundance gives a good fit.

\subsection{Jump models}
\label{sec:mod_jump}

\begin{figure}[t]
  \begin{center}
    
\psfig{file=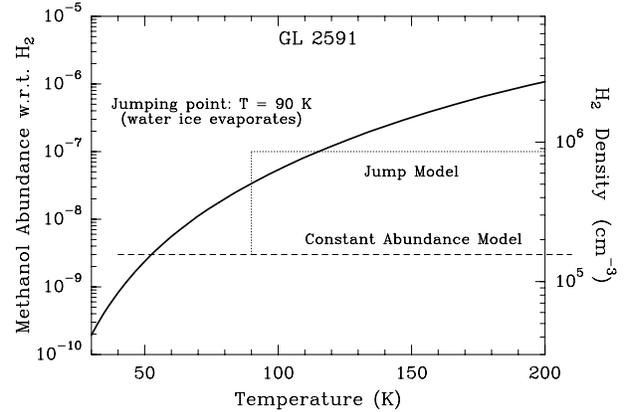,width=9cm,angle=-90}

    \caption{Density and temperature structure (solid line) and
      abundance profiles (dotted and dashed lines) of the two models
      for \meth\ in the source GL 2591.}
    \label{f:jump}

  \end{center}
\end{figure}

For the sources W~33A, GL~2591 and NGC~7538~IRS1, the lines from
energy levels $\gtsim 100$~K above ground require significantly higher
abundances than the lower-excitation lines.  These results suggest
that the abundance of \meth\ is higher in the warm, dense gas close to
the star than in the more extended, cold and tenuous gas. From the
location of the break in the abundance profile, it seems likely that
evaporation of grain mantles, which also occurs at $\sim 100$~K, plays
a role. As a simple test of ice evaporation, we have run models for
these three sources where the \meth\ abundance follows a step
function. In these ``jump'' models, the \meth\ abundance is at a low
level, the ``base level'', far from the star, while at a threshold
temperature, the abundance surges to a high value, the ``top level''.
The situation is sketched in Figure~\ref{f:jump}.  The motivation for
this model is that if methanol is present in icy grain mantles, its
abundance will increase strongly when these ices evaporate. For the
temperature threshold, we take $90$~K, which is where water ice, the
most refractory and most abundant component of the grain mantles,
evaporates in $\sim 10$~yr (\cite{sand93}).

Alternatively, the ice may be desorbed in a weak shock associated with
the molecular outflows these sources are known to have: \cite{ant96}
calculate that a local grain-grain velocity dispersion of $\approx
2$~\kms\ is sufficient to shatter ice material.  Shock desorption of
methanol is known to be important in low-mass protostars, for example
L~1157 (\cite{bach95}) and NGC~1333 IRAS~4A (\cite{blak95}).  The
large line widths and line profiles measured here for the
low-luminosity sources GL~7009S and IRAS 20126 indicate that shock
desorption plays a role there; for the other sources, thermal effects
probably dominate.  Since \meth\ is confined to the refractory (polar)
component of the ice, it is not necessary to consider the effect of a
slowly rising temperature in the hot core region, as \cite{viti99}
did.

For the base level, we take the abundances found in the previous
section, which were constrained mostly by the low-$K$ lines with
$E<100$~K.  We have considered jump factors of 3, 10, 30, 100 and 300
without further iteration. Between these models, jumps by factors of
$\sim 30$ give the best match to the high-excitation lines.  The
results of these models are plotted as squares in
Fig.~\ref{fig:const_x} and listed in column~6 of Table~\ref{t:trot}.

\begin{figure}[b]
  \begin{center}
    
\psfig{file=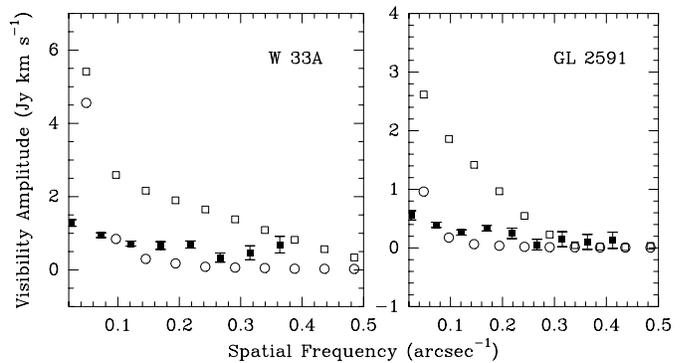,width=9cm,angle=-90}

    \caption{Visibility amplitudes of \meth\ $3_1 \to 4_0$~A$^+$ emission as observed
      with OVRO toward W~33A \textit{(left)} and GL~2591
      \textit{(right)}, and model points for the constant-abundance
      \textit{(open circles)} and the ``jump'' models \textit{(open
        squares)}.}
    \label{f:m_ovro}

  \end{center}
\end{figure}

Further constraints on the abundance profile of \meth\ in our sources
may be obtained by comparing the models to the OVRO data.  The points
in Fig.~\ref{f:m_ovro} are the OVRO visibility data of W~33A and
GL~2591. The observations of the $107$~GHz line have been integrated
over velocity and binned in annuli around the source position.
Superposed are model points for the constant-abundance model and for
the jump model. The data do not particularly favour one model or the
other, perhaps because this transition traces mostly cold
gas. Collisional rate coefficients up to $J=11$ are eagerly awaited,
so that the $104$~GHz line can be modeled as well.

\section{Discussion}
\label{sec:disc}

\subsection{Relation to solid-state observations}
\label{sec:solid}

The enhancement of the methanol abundance at high temperatures is very
likely related to the evaporation of solid methanol, which has been
observed at infrared wavelengths toward many of the sources studied
here (\S~\ref{sec:intro}). Ground-based data already revealed that
\meth\ is not mixed with the bulk of the water ice (\cite{skin92}).
Recent ISO results indicate extensive ice segregation toward massive
young stars; the absorption profiles of \coo\ ice indicate that the
methanol resides in a \water:\meth:\coo =1:1:1 layer heated to
$\approx 80$~K as shown by \cite{pasc98}, \cite{perry99} and
\cite{dart99a}.  The column density of solid \meth\ is typically $\sim
5$\% of that of water ice, but in two cases, W~33A and GL~7009S, it is
as high as $30$\%. It has been suggested that \meth\ ice is
preferentially formed around more massive young stars, but among the
sources studied here, these two do not stand out in luminosity.
Instead, they are the two most embedded sources in the sample (see van
der Tak et al.\ 2000\nocite{fvdt00}), which may enhance the formation
of solid methanol. The gas-phase abundance of \meth\ 
(Table~\ref{t:trot}) is much lower in GL~7009S than that in W~33A,
suggesting that GL~7009S is in an earlier evolutionary state where ice
evaporation affects only a small part of the envelope.

A link between solid and gaseous methanol is plausible because the
sources with high gas-phase \meth\ abundances are also the ones with
high fractions of annealed solid $^{13}$\coo\ (\cite{boog00}), and
high abundances of gas-phase \water\ and \coo\ (Boonman et al.\ 
2000)\nocite{boon00}.  These molecules evaporate at $\sim 90$~K, like
methanol.  The \meth\ abundances also follow the ratios of envelope
mass to stellar mass and the 45/100~\mic\ colours from \cite{fvdt00},
which confirms the picture that warmer sources have higher molecular
abundances in the gas phase.  These results indicate that the
excitation and abundance of gaseous \meth\ can be used as evolutionary
indicators during the embedded stage of massive star formation. As
discussed in \S~\ref{sec:res_jcmt}, using submillimeter data to trace
evolution has the advantage of being independent of source orientation
or total mass, because the dust emission is optically thin at these
wavelengths.

Grain mantle evaporation appears to be much less important for \hhco\ 
than for \meth.  We have compared the \hhco\ data to the
constant-abundance models from \cite{fvdt00} in similar plots as
Figure~\ref{fig:const_x}, and found good agreement. There is no
evidence for jumps in the \hhco\ abundance by factors $\gtsim 3$
within the temperature range of $20-250$~K that the observations are
sensitive to.  This result suggests that the formaldehyde observed in
these sources is predominantly formed in the gas phase by oxidation of
CH$_3$. Gas-phase models by \cite{lee96} indicate an abundance of
$\sim 10^{-9}$, similar to the observed value, although a contribution
from ice evaporation at the $10^{-9}$ level cannot be excluded. 

The high observed HDCO/\hhco\ ratios and the detections of
  D$_2$CO toward the Compact Ridge in Orion (Wright et al.\ 1996;
  Turner 1990) \nocite{melw96,turn90,evd95,cecc98} and toward embedded
  low-mass objects such as IRAS 16293 (van Dishoeck et al.\ 1995;
  Ceccarelli et al.\ 1998) indicate \hhco\ formation on grains.  In
  our sources, HDCO is not detected to \hhco/HDCO $>10$; in a survey
  of hot core-type sources, \cite{jenn98} obtained DCN/HCN $\sim
  10^{-3}$, much lower than in embedded low-mass stars. Although the
  surface chemistry should qualitatively be the same, the cold ($\sim
  10$~K) phase may last too short in our sources to build up large
  amounts of deuterated molecules such as those seen in the Compact
  Ridge and in low-mass objects.
  
  In eleven regions of (mostly massive) star formation, \cite{mang93}
  derived ratios of ortho- to para-\hhco\ of $1.5-3$ and took this
  result as evidence for grain surface formation of \hhco.  However,
  this conclusion appears tentative, since for most of their sources,
  only one line of ortho-\hhco\ was observed. In addition, at the
  \hhco\ column densities of $\sim 10^{14}$~\scm, representative of
  the regions studied by Mangum \& Wootten and by us, the $K=0$ and
  $K=1$ lines of \hhco\ have optical depths of $\sim 1$, requiring
  careful modeling.  The data and models presented in this paper are
  consistent with an ortho-/para-\hhco\ ratio of~3.  

\bigskip

High abundances of methanol as observed in the ices, $\sim 10^{-6}$
relative to hydrogen, cannot be produced in the gas phase, except
maybe in shocks. This mechanism has been proposed for \water\ ice by
\cite{berg99} and for \coo\ by \cite{char00}; its application to
\meth\ depends on the existence of a high-temperature route to form
methanol, which is not yet known.  \cite{hart95} proposed the reaction
CH$_4+$OH, but ISO-SWS observations by \cite{boog98} indicate low
abundances of gaseous methane in our sources, and at $T\gtsim 200$~K,
all OH should be consumed by H and \hh\ to form \water. The widths of
the \meth\ lines of only a few \kms\ also argue against formation
in shocks.


More likely, the solid methanol is formed by reactions on or inside
the ice layers.  Addition of H~atoms to CO molecules will lead to
\hhco, and further to \meth. In the literature, there are three
proposed sources of atoms: direct accretion from the gas phase,
ultraviolet irradiation and bombardment by energetic particles. We
show in Section~\ref{sec:irr} that the latter two mechanisms are
unlikely to be important for the sources studied in this paper, and
focus for now on the first.

\subsection{Surface chemistry model}
\label{sec:xander}

If the \meth\ ice that we see evaporating in these sources originates
from H~atom addition (=reduction) of CO ice on the surfaces of dust
grains, it cannot have been produced under the current physical
conditions. The evaporation temperature of \meth\ and \water\ ice,
$90$~K, is much higher than that of CO and O, $\approx 20$~K, and that
of H, $\approx 10-15$~K, depending on its surface mobility and
reactivity.  Any formation of \meth\ ice through surface reduction of
CO must therefore have occurred before the central star heated up its
envelope above $\sim 15$~K. The most likely phase of the cloud to form
methanol ice through surface chemistry is therefore the
pre-protostellar phase, when the grain temperature may have been as
low as $\approx 10$~K and the cloud was contracting to form a dense
core. However, about 1/3 of solid CO is observed inside the water ice
layer, and will not evaporate until much higher temperatures are
reached (\cite{tiel91}). If H and O atoms can be similarly trapped,
solid-state reduction and oxidation may occur at temperatures well
above $20$~K.

Tielens \& Hagen (1982) and Tielens \& Allamandola (1987)
\nocite{tiel82,tiel87} proposed that direct accretion of atoms and
molecules from the gas phase, followed by low-temperature surface
reactions, determine the composition of grain mantles.  Activation
barriers are offset by the long effective duration of the collision on
the surface. The resulting mantle composition is determined by the
relative accretion rates of H~and CO onto the grains, and the relative
height of the reaction barriers.

\begin{figure}[t]
\begin{center}

\psfig{file=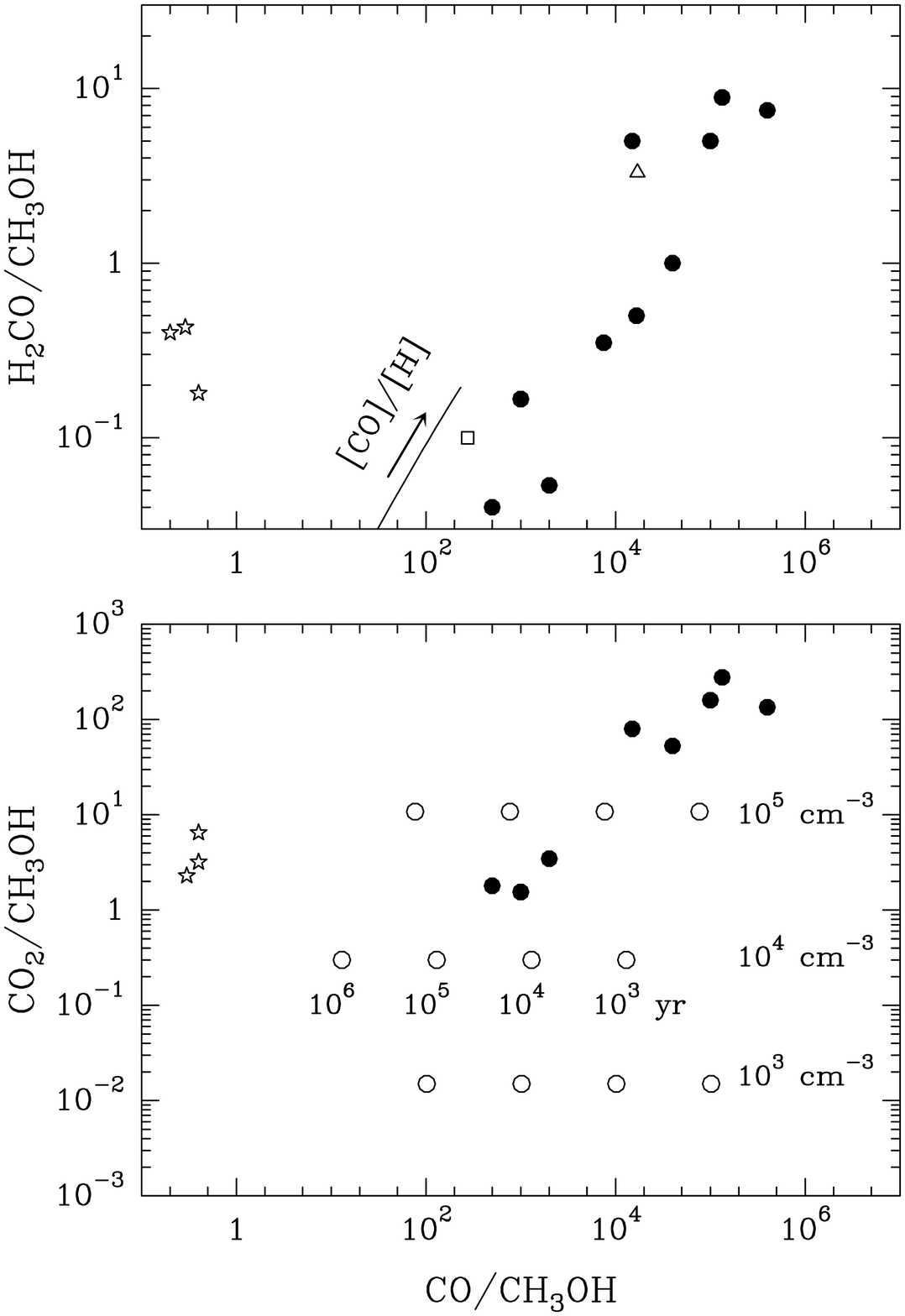,width=9cm}

\caption{ Filled circles: observed abundance ratios of \hhco/\meth\ (top) and
  \coo/\meth\ (bottom) versus that of CO/\meth. Open triangle: data on
  translucent clouds from \cite{turn98}; open square: data on the
  compact ridge in Orion from \cite{sutt95}. Stars: solid-state data
  from \cite{jacq00}. Heavy line: surface chemistry model after
  \cite{ctr97}.  Open circles: results of surface chemistry model,
  labeled by time and density of the gas phase.}
\label{f:ctd}

\end{center}
\end{figure}

This mechanism predicts that the abundance ratios of \hhco/\meth\ and
CO/\meth\ are correlated, as shown by the curve in Fig.~\ref{f:ctd},
taken from \cite{ctr97}, with the reaction probability ratio
$\phi_{\rm H}$ of CO + H to \hhco\ + H is $10^{-3}$. Along the curve,
the H/CO abundance ratio decreases down to $10^{-2}$ at its tip.  At
the density in the central regions of our sources, $\sim 10^6$~\ccm,
atomic~H is mainly produced by cosmic-ray ionization of \hh, giving a
constant concentration of $\sim 1$~\ccm, or an abundance of $10^{-6}$.
The abundance of CO in these sources was measured by \cite{fvdt00} to
be $\approx 1\times 10^{-4}$, as expected if all gas-phase carbon is
in CO and no significant depletion occurs. Hence, H/CO $\approx
10^{-2}$, and our sources should lie right at the tip of the curve.
The value $\phi_{\rm H}=10^{-3}$ implies that the CO$\to$ \hhco\ 
reaction limits the rate of \meth\ formation, consistent with the
non-detection of evaporated \hhco.

The figure also plots the observed abundance ratios, where the \hhco\ 
and CO abundances have been taken from \cite{fvdt00}, while for \meth,
either the constant-abundance model or the ``top level'' of the jump
model has been plotted.  The data points are seen to lie along a
``sequence'' which lies along the model curve, perhaps displaced
toward lower \meth\ abundances by an order of magnitude. Although this
behaviour is consistent with solid-state formation of \hhco\ and
\meth\ at various H/CO ratios, the different rotation temperatures and
abundance profiles of \hhco\ and \meth\ argue against a common
chemical origin for these two species, and favour complete
hydrogenation of CO into \meth. The figure also plots solid state
observations by \cite{jacq00}, which are lower limits in the case of
CO because of evaporation.

Since the abundances of CO and \hhco\ are almost the same in all our
sources, the spread in the abundance ratios is probably due to
different degrees of \meth\ evaporation and subsequent gas-phase
destruction. Indeed, the abundances of gaseous \meth\ are factors of
$\gtsim 10$ lower than the values measured in the solid state. This
difference is expected since methanol is destroyed in $\sim 10^4$~yr
(Charnley et al.\ 1992)\nocite{char92}
in reactions with ions such H$_3^+$, leading to a rich chemistry with
species like CH$_3$OCH$_3$ and CH$_3$OCHO.  These species are indeed
detected toward the ``hot cores'' W~3 (\water) and NGC~6334 IRS1,
weaker toward the ``jump sources'' GL~2591 and NGC~7538 IRS1, and
absent in the low-methanol source W~3 IRS5 (Helmich \& van Dishoeck
1997; van Dishoeck \& van der Tak 2000)\nocite{helm97,korea00}. 

We conclude that the observed spread in the gaseous methanol abundance
is due to incomplete evaporation for the cold sources, which are the
least evolved ones in our sample, and to gas-phase reactions for the
hot cores, which are the most evolved. The most likely source of the
methanol is grain surface chemistry in the pre-protostellar phase. The
conditions in this phase cannot be derived by comparison to CO and
\hhco, for which our data suggest that gas-phase processes control the
chemistry.  A more promising molecule is \coo.

\subsection{The H/O competition: Density and duration of the
  pre-stellar phase}
\label{sec:paola}

The ISO detection of ubiquitous solid \coo\ by \cite{perry99} makes
clear that oxidation of CO is a potentially important competitor to
reduction. These data are shown in the bottom panel of
Fig.~\ref{f:ctd}, as well as gas-phase \coo\ abundances from
\cite{boon00}.  Gas-grain chemistry is clearly important for \coo, as
it is for \meth, warranting a comparative study of the two.  The model
by \cite{ctr97} did not include reactions with O, while \cite{tiel82}
used \oo\ as the dominant form of oxygen, which is inconsistent with
recent observational limits (\cite{meln00}). We have constructed a
model of gas-grain chemistry based on the modified rate equation
approach described in \cite{casel98}, and extensively tested against a
Monte Carlo program.  The chemical system consists of three gaseous
species: H, O~and CO, and eight surface species: OH, \water, \oo,
\coo, HCO, \hhco, CH$_2$OH and \meth. The calculations assume fast
quantum tunneling of hydrogen, as proposed by \cite{tiel82}. If
hydrogen does not tunnel, as measurements by \cite{katz99} suggest,
our simple system is unable to form significant mole fractions of
\meth\ and \coo. In the case of reduced mobility for all species, the
Eley-Rideal mechanism should be considered to model surface chemistry
(\cite{eric00}), which is not included here.

The mantle composition (Table~\ref{t:paola}) depends on density
through the composition of the gas phase, in particular the H/O ratio.
We take the concentrations of H, O and CO from \cite{lee96}, for the
case of low metals, $T=10$~K, in steady state. The main result is that
after reduction to HCO, CO is mostly reduced into \meth\ at low
densities and mostly oxidized at higher densities. In our model, the
reaction CO+O$\to$\coo\ has a barrier of 1000~K (see \cite{dhen85}),
else ices of 60\% \coo\ would form which are not observed. Water is
copiously made through the sequence O+H$\to$OH and OH+H$\to$\water.
The detailed composition of the grain mantles is quite uncertain
because of unknown reaction rates. For instance, the fraction of
\hhco\ depends on relative barrier height of the CO+H and \hhco+H
reactions, but \hhco\ is never a major component of the ice layer,
consistent with gas-phase and solid state observations. The results in
Table~\ref{t:paola} refer to a chemical time scale of 1000 years.
Unlike inert species such as \oo, \water, \meth\ and \coo, the
fractions of trace species (fractions $\ltsim 0.01$) also depend
somewhat on time, and are not used in the analysis. Species that react
quickly with H maintain a constant population because they are
hydrogenated on the H accretion time scale ($\sim 1$~day if $n_{\rm
  H}=10^3$~\ccm), and their molar fractions decline as the ice layer
grows with time.

As an example, we will now estimate the yield of \meth\ by this scheme
for the case $n_{\rm H} = 10^4$~\ccm. The depletion rate of CO molecules 
from the gas phase is $f_d = n_d \sigma V_{\rm CO} S$, with $n_d$ and
$\sigma$ the number density and cross section of the dust grains,
$V_{CO}$ the thermal velocity of CO and $S$ the sticking coefficient,
assumed to be $10$\%. Taking silicate grains of radius $0.1$~\mic,
$f_d = 3 \times 10^{-15}$~s$^{-1}$, or $10$\% in $10^6$ years.
At this density, $77$\% of this depleted CO goes into \meth,
so that upon evaporation, the abundance ratio would be \meth/CO
$\approx 0.086$ or \meth/\hh $\sim 10^{-5}$. 

\begin{table}[t]
\caption{Results of gas-grain chemical model at $T=10$~K.}
  \begin{center}
    \begin{tabular}{llll}

      \hline
Species & \multicolumn{3}{c}{Total density $n_{\rm H}$} \\
        & $10^3$ & $10^4$ & $10^5$ \\ 
\multicolumn{4}{c}{\hrulefill} \\
\multicolumn{4}{c}{Gas phase concentrations (\ccm)} \\
\multicolumn{4}{c}{\hrulefill} \\
H  & 1.15 & 1.15 & 1.1 \\
O  & 0.09 & 0.75 & 7.0 \\
CO & 0.075 & 0.75 & 7.5 \\
\multicolumn{4}{c}{\hrulefill} \\
\multicolumn{4}{c}{Molar fractions in the solid state} \\
\multicolumn{4}{c}{\hrulefill} \\
\oo\    & 9.3(-3) & 0.11    & 0.27 \\
\water\ & 0.60    & 0.37    & 0.12 \\
\hh\    & 0.0     & 0.0     & 0.0 \\
CO      & 1.4(-5) & 3.9(-5) & 0.48 \\
\hhco\  & 1.4(-5) & 3.1(-5) & 3.1(-2) \\
\meth\  & 0.39    & 0.41    & 7.8(-3) \\
\coo\   & 6.1(-3) & 0.12    & 8.7(-2) \\
\multicolumn{4}{c}{\hrulefill} \\

\end{tabular}
\end{center}
\label{t:paola}
\end{table}

Assuming that the composition of ice layers is determined by surface
chemistry, our model can be used to investigate the initial conditions
of massive star formation by considering abundance ratios of CO, \coo\ 
and \meth.  The \coo/\meth\ ratio is sensitive to density, independent
of time, and equals the ratio of the molar fractions of solid \coo\ 
and \meth\ in Table~\ref{t:paola}.  Time is measured by the ratio of
raw material (CO) to product (\meth) through the fraction of CO in the
solid state derived above, $0.01 (t/10^5 {\rm yr}) (n_{\rm H} / 10^4
{\rm cm}^{-3})$. The abundance ratio in the gas phase is this fraction
multiplied by the efficiency of CO$\to$\meth\ conversion, equal to the
molar fraction of \meth\ divided by the sum of the fractions of CO and
its possible products, \coo, \hhco\ and \meth.  Figure~\ref{f:ctd}
plots the synthetic abundance ratios as open circles, labeled by time
and density. If this model is valid, the observed \coo/\meth\ ratio
constrains the density in the pre-protostellar phase to be $\approx
10^5$~\ccm, or higher, since some \coo\ may be destroyed shortly after
evaporation. At this density, hydrogenation of CO is incomplete due to
the low H flux, and a significant abundance of solid \hhco\ is
expected, consistent with the results of \cite{jacq00}. Using this
limit on the density, the observed CO/\meth\ ratio constrains the time
spent in the pre-protostellar phase to be $\ltsim 10^5$~years within
the grain surface chemistry scenario. The same conclusion is reached
when the CO/\coo\ ratio is used instead of CO/\meth; the uncertainty
is a factor of ten due to the unknown sticking coefficient.  This time
scale is significantly smaller than the corresponding number for
low-mass stars, where this phase lasts $\sim 10^6$~years as derived
from the ratio of dense cores with and without stars found by
\cite{beich86}. The same time scale of $\sim 10^6$~yr is obtained from
our models for an abundance ratio of CO/\meth\ $\sim 10^3$, as
observed in the compact ridge in Orion.

These models also help to understand observations of solid \coo\ in
other regions where no star formation is occurring.  Toward the field
star Elias~16 behind the Taurus dark cloud, \coo\ has been detected
(\cite{whit98}) but \meth\ has not (\cite{chiar98}). The models
suggest thus that the density in this region is $\gtsim 3\times
10^4$~\ccm. The model does not explain the lack of solid \meth\ 
toward SgrA$^*$ ($\ltsim 3$\% relative to \water\ ice;
\cite{chiar00}), where \coo\ ice has been detected. For this
low-density line of sight, the opposite ratio would be expected. An
enhanced temperature or ultraviolet irradiation may be important here.
For the disk around NGC~7538 IRS1, where $N$(\hhco)/$N$(\meth) $\gtsim
1$ (\S~\ref{sec:maser}), the models indicate that the density is
$\gtsim 10^5$~\ccm.

The model is also consistent with current observational limits on
solid \oo.  At low densities, our models drive all O into \water\ on
the dust grains, but for $\ge 10^5$~\ccm, part of it goes into \oo.
However, the ratio \oo/CO remains $<1$ on the surface, in agreement
with the limit derived by \cite{buss99} from ISO-SWS observations of
NGC~7538 IRS9. Evaporation of solid \oo\ plays a role, but observations
of gaseous \oo\ with the Submillimeter Wave Astronomy Satellite give
an upper limit toward massive star-forming regions of $\sim 10^{-7}$
relative to \hh\ (\cite{meln00}), consistent with the solid-state
results. 

\subsection{Alternative models}
\label{sec:irr}

Could ultraviolet irradiation or energetic particle bombardment also
produce the observed trends in the abundances? The production rate of
species $i$ by irradiation can be written as $dn_i/dt = \alpha_i \Phi
4\pi r_g^2$, with $\Phi$ the ultraviolet flux in photons \scm~s$^{-1}$
and $r_g$ the radius of the dust grains, taken to be $10^{-5}$~cm.
Inside dense clouds, most ultraviolet radiation is produced by
cosmic-ray interaction with \hh, which gives a field approximately
equal to the interstellar radiation field at $A_V=5$ (\cite{pras83}),
or $5000$ photons \scm~s$^{-1}$. The reaction yields $\alpha_i$ follow
from laboratory experiments such as those reported at the Leiden
Observatory Laboratory database\footnote{see
  www.strw.leidenuniv.nl/$\sim$lab}, in this case on a mixture of
initial composition \water:CO = 100:33. The amounts of \meth\ and
\coo\ produced do not depend on this ratio as long as it is $>1$.
Using band strengths by \cite{perry96} and \cite{oswin99}, we obtain
$\alpha=1.9 \times 10^{-3}$ for \meth\ and $\alpha=3.2 \times 10^{-2}$
for \coo. The ratio of these numbers is in good agreement with the
observed abundance ratio of $\sim 10$. However, the absolute values of
the $\alpha_i$ imply production rates of $\sim 10^{-8}$~s$^{-1}$ per
grain, compared to the accretion rate of H, O and CO of $\sim
10^{-5}$~s$^{-1}$. Stellar ultraviolet radiation can only affect the
inner parts of the envelopes since their extinctions are $A_V \sim
100$ magnitudes. The absence of a strong ultraviolet field throughout
the envelopes of our sources is also suggested by the observational
limits from ISO-SWS on mid-infrared fine structure lines and of
emission by polycyclic aromatic hydrocarbons (\cite{korea00} 2000).

The processing of interstellar ice by cosmic rays was studied by
\cite{huds99}, who bombarded an \water:CO=5:1 ice mixture with protons
of energy $\approx 0.8$~MeV. These experiments produced abundance
ratios of \hhco/\meth=0.6 and \coo/\meth=2.0, in fair agreement with
the observed values. Other species formed as well, notably formic acid
(HCOOH) and methane (CH$_4$), which indeed are observed in
interstellar grain mantles. However, the experiments by Hudson \&
Moore produced almost twice as much HCOOH and CH$_4$ as \coo, while
the observed abundances are only $\approx 10$\% of that of \coo\ 
(\cite{willem99}; Boogert et al.\ 1998\nocite{boog98}). In addition,
the particle dose in the experiments was $2\times 10^{15}$~\scm, while
the interstellar cosmic-ray flux is only 3~\scm~s$^{-1}$. Hence, the
experiments simulate a bombardment for $\sim 3\times 10^7$~yr, a
factor of $\sim 1000$ longer than the ages of the sources studied
here. Stellar X-ray emission (\cite{glass00}) only acts on small scales,
especially because of heavy attenuation in the envelopes of these
obscured objects.

\section{Conclusions}
\label{sec:conc}

The chemistry of \meth\ and \hhco\ in thirteen regions of massive star
formation is studied through single-dish (JCMT) and interferometer
(OVRO) line observations at submillimeter wavelengths. Our main
conclusions are:

1. The submillimeter emission lines of \meth\ toward most sources have
widths of $3-5$~\kms, consistent with those found earlier for \co\ 
and \cs. However, in the low-luminosity sources GL~7009S and IRAS 20126,
the line shapes reveal that \meth\ is present in the outflow.  These
results indicate that the desorption of ices in the envelopes of
low-mass protostars is primarily by shocks, while thermal processes
dominate in the case of massive stars.

2.  Rotational temperatures of \meth\ range from $10$ to $200$~K and
correlate very well with the excitation temperature of \hcch\ measured
in infrared absorption. This correlation suggests that both species
trace the same gas, which may be outgassing of icy grain mantles. For
\hhco, the range in $T_{\rm rot}$ is only $60-90$~K without relation
to $T_{\rm ex}$(\hcch), suggesting a different chemical origin. We
propose that $T_{\rm rot}$ (\meth) can be used as evolutionary
indicator during the embedded phase of massive star formation,
independent of source optical depth or orientation.
  
3. Detailed non-LTE radiative transfer models of the \meth\ lines
suggest a distinction of three types of sources: those with
\meth/\hh$\sim 10^{-9}$, those with \meth/\hh$\sim 10^{-7}$ and those
which require a jump in its abundance from $\sim 10^{-9}$ to $\sim
10^{-7}$.  The models are consistent with the $\approx 3''$ size of
the \meth\ $107$~GHz emission measured interferometrically. The
location of the jump at $T\approx 100$~K strongly suggests that the
methanol enhancement is due to evaporation of icy grain mantles. The
sequence of low-methanol $\to$ jump $\to$ high-methanol sources
corresponds to a progression in the ratio of envelope mass to stellar
mass and the mean temperature of the envelope.  In contrast, the
observed \hhco\ 
seems primarily produced in the gas phase, since the \hhco\ data can
be well fit with a constant abundance of a few $\times 10^{-9}$
throughout the envelope. The grain surface hydrogenation of CO thus
appears to be completed into \meth, with little \hhco\ left over.

4. Model calculations of gas-grain chemistry show that CO is primarily
reduced (into \meth) at densities $n_{\rm H} \ltsim 10^4$~\ccm, and
primarily oxydized (into \coo) at higher densities.  To keep
sufficient CO on the grains, this mechanism requires temperatures of
$\ltsim 15$~K, i.e., conditions before star formation.  Assuming that
surface reactions proceed at the accretion rate of CO, the observed
\coo\ and \meth\ abundances constrain the density in the
pre-protostellar phase to be $n_{\rm H} \gtsim$ a few $10^4$~\ccm, and
the time spent in this phase to be $\ltsim 10^5$~yr. Our surface
  chemistry model predicts that lines of sight through clouds with a
  high H/O ratio will show abundant solid methanol and less \coo.
Ultraviolet photolysis and radiolysis by energetic (MeV) protons
appear less efficient as ice processing mechanisms for these sources;
radiolysis also overproduces HCOOH and CH$_4$.

\begin{acknowledgements} 
  
  The authors wish to thank Malcolm Walmsley for providing collisional
  rate coefficients for methanol, Xander Tielens, Pascale Ehrenfreund
  and Willem Schutte for comments on the manuscript and Ted Bergin,
  Eric Herbst, Friedrich Wyrowski and Wilfried Boland for useful
  discussions. Annemieke Boonman and Jacquie Keane kindly provided us
  with their results in advance of publication. We are grateful to
  Remo Tilanus and Fred Baas at the JCMT and Geoffrey Blake at OVRO
  for assistance with the observations. This research was supported by
  NWO grant 614-41-003 and the MURST program ``Dust and Molecules in
  Astrophysical Environments''.

\end{acknowledgements} 

\bibliographystyle{astron}

\end{document}